\documentclass[12pt,a4paper]{article}

\usepackage[dvips]{graphicx}
\usepackage{amsmath}
\newcommand{\dtilde}[1]{ \tilde{\tilde{#1}} }
\newcommand{\be}{\begin{eqnarray} }
\newcommand{\ee}{\end{eqnarray} }

\begin{document}
 \begin{center}
         {\large \bf
         Valence and sea contributions to the nucleon spin
         }
 \end{center}
 \begin{center}
 \vskip 1cm
 {\large O.\,Yu.\,Shevchenko}\footnote{E-mail address: shev@mail.cern.ch},
 {\large R.\,R.\,Akhunzyanov}\footnote{E-mail address: axruslan@mail.ru},
 {\large V.\,Yu.\,Lavrentyev}\footnote{E-mail address: basil77@jinr.ru}
 
 \vspace{0.5cm}
 {\it Joint Institute for Nuclear Research}
 \end{center}

%\title{Valence and sea contributions to the nucleon spin}
%\author{
%O.\,Yu.\,Shevchenko\/\thanks{e-mail: shev@mail.cern.ch},
%R.\,R.\,Akhunzyanov\/\thanks{e-mail: axruslan@mail.ru},
%V.\,Yu.\,Lavrentyev\/\thanks{e-mail: basil77@jinr.ru}
%}
%\institute{Joint Institute for Nuclear Research}

\begin{abstract}
The first moments of the polarized valence parton distribution functions (PDFs) 
truncated to the wide Bjorken $x$ region $0.004<x<0.7$
are directly (without any fitting procedure) extracted in 
the next to leading order (NLO) QCD
from both 
COMPASS and HERMES data 
on pion production in polarized semi-inclusive DIS (SIDIS) experiments. 
The COMPASS and HERMES data are combined in two ways and 
two scenarios for the fragmentation functions (FFs) are considered.
Two procedures are proposed for an estimation of light sea quark contributions 
 to the proton spin.
%to estimate the sea contributions to the proton spin are proposed.
Both of them lead to the conclusion that these contributions are compatible 
with zero within the errors.
\end{abstract}
\begin{flushleft}
\end{flushleft} 
%\PACS{
%      {13.65.Ni}{}   \and
%      {13.60.Hb}{}   \and
%      {13.88.+e}{}
%     } % end of PACS codes
%} 
%end of abstract

%PACS: 13.65.Ni, 13.60.Hb, 13.88.+e

%\maketitle

Longitudinally polarized parton distribution functions (PDFs), and especially
their first moments, which directly compose the nucleon spin together with the orbital parton
momenta, are of crucial importance for solution of the proton spin puzzle, attracting
great both theoretical and experimental efforts during many years. 
Nowadays, there is a huge growth of interest to the semi-inclusive DIS (SIDIS) experiments with
longitudinally polarized beam and target such as SMC \cite{smc}, HERMES \cite{hermes}, 
COMPASS \cite{compass} (see, for instance, \cite{our-review} for review).
It is of importance that the SIDIS experiments, where one identifies the hadron in the 
final state,
provide us with the additional information on the partonic spin structure in comparison with
the usual DIS experiments. Namely, in contrast to the DIS data,  
the SIDIS data allows us to find the sea and valence 
contributions to the nucleon spin in separation. In this paper we just focus on this important
task. 
To this end we apply 
the original procedure 
of the polarized SIDIS data analysis in NLO QCD, 
elaborated in the sequel of papers \cite{prd2004,jetp2005,prd2006} 
(see also \cite{our-review} for more details).
It is of importance (especially for analysis of still 
relatively poor SIDIS data we deal with)
that this alternative procedure allows to extract the truncated to the 
accessible for measurement Bjorken $x$ region moments of polarized
PDFs directly, without any fitting procedure 
(with unavoidable arbitrariness in the choice of functional form of $\Delta q$  
at initial $Q^2$ and a lot of free varying parameters): 
within this procedure the central values of asymmetries and their uncertainties 
directly propagate
to the extracted values of moments and their errors.
Of importance also that we use in our analysis the difference asymmetries
(just as in Refs. \cite{prd2004,prd2006}),
that allows
to avoid the application of poorly known 
%fragmentation functions (FFs)
FFs (such as $D_q^{K^\pm}$ and $D_g^h$ FFs) -- 
see \cite{prd2004,prd2006} and also \cite{our-review} for review.

Within this paper we use two NLO parametrizations  AKK08 \cite{akk08} and DSS \cite{dss} 
for FFs, which differ quite essentially in the pion sector.
AKK08 parametrization corresponds to unbroken $SU_f(2)$ symmetry:
\begin{equation}
\label{D_fav}
D_1 \equiv D_{u}^{\pi^+} \stackrel{C}{=} D_{\bar{u}}^{\pi^-}
\stackrel{SU(2)}{=} D_{\bar{d}}^{\pi^+}
\stackrel{C}{=} D_{d}^{\pi^-},
\end{equation}
\begin{equation}
\label{D_unfav}
D_2 \equiv D_{\bar{u}}^{\pi^+} 
\stackrel{C}{=} D_{u}^{\pi^-}
\stackrel{SU(2)}{=} D_{d}^{\pi^+}
\stackrel{C}{=} D_{\bar d}^{\pi^-}.
\end{equation}
On the other hand, DSS parametrization \cite{dss} allows violation of $SU_f(2)$ symmetry
in the sector of the favored pion FFs:
\begin{equation}
\label{dssD1}
D_1\equiv D_{u}^{\pi^+} 
\stackrel{C}{=} D_{\bar{u}}^{\pi^-},
\end{equation}
\begin{equation}
\label{dssD11}
\tilde{D_1} \equiv D_{\bar d}^{\pi^+}
\stackrel{C}{=} D_{d}^{\pi^-},
\end{equation}
\begin{equation}
\label{dssD2}
D_2 \equiv D_{\bar u}^{\pi^+} 
\stackrel{C}{=} D_{u}^{\pi^-}
\stackrel{SU(2)}{=} D_{d}^{\pi^+}
\stackrel{C}{=} D_{\bar d}^{\pi^-},
\end{equation}
so that the favored FFs $D_1$ and $\tilde{D_1}$ of $u$ and $d$ quarks are not equal 
to each other.

For the scenario preserving $SU_f(2)$ symmetry
(AKK08 parametrization for FFs here)
the procedure of direct extraction in NLO QCD of {\it n}-th
moments of the valence PDFs from the measured difference asymmetries
is described in Refs. \cite{prd2004}, \cite{prd2006} in details. 
Let us recall the key necessary equations.

The theoretical expressions\footnote{
Since namely $F_2$ function is measured in experiment and then is used
to parameterize unpolarized PDFs, 
the form of these equations used here
is the most convenient to properly account for correction due to 
the factor $R = \sigma_L/\sigma_T$ (see discussion around Eq. (10) in \cite{compass2007},
around Eq. (12) in \cite{EPJ2010} and references therein).
}  
for the difference asymmetries in NLO QCD look as
\begin{equation}
%\begin{multline}
\label{apdif2}
A_{p}^{\pi^+-\pi^-}(x,Q^2){\Bigl |}_Z  = (1+R)\frac{
\int_Z^1 dz_h (4\Delta u_V-\Delta d_V) \Delta \hat{K} (D_1-D_2)}
{\int_Z^1 dz_h (4u_V-d_V) \hat{K} (D_1-D_2)},
%\end{multline}
\end{equation}
for the proton target,
and
\begin{equation}
%\begin{multline}
\label{addif2}
A_{d}^{\pi^+-\pi^-}(x,Q^2){\Bigl |}_Z  = (1+R)(1-\frac{3}{2}\omega_D)\frac{ 
\int_Z^1 dz_h (\Delta u_V+\Delta d_V) \Delta \hat{K} (D_1-D_2)}
{\int_Z^1 dz_h (u_V+d_V) \hat{K} (D_1-D_2)},
%\end{multline}
\end{equation}
for the deuteron target.
Here, for brevity, we have introduced the operator notation
$\hat K$ and $\Delta \hat K$:
\begin{equation}
\hat{K} \equiv 1 + \otimes \frac{\alpha_s}{2\pi} C^2_{qq} \otimes, \quad 
\Delta \hat{K} \equiv 1 + \otimes \frac{\alpha_s}{2\pi} \Delta C_{qq} \otimes,
\end{equation}
so that
\begin{eqnarray}
%\begin{multline}
[q\, \hat{K}\, D](x,z_h) =
q(x) D(z_h)  +
\frac{\alpha_s}{2\pi} \int_x^1 \frac{dx'}{x'} \int_{z_h}^1 \frac{dz'}{z'} 
q \left(\frac{x}{x'}\right) C^2_{qq}(x',z') D\left(\frac{z_h}{z'}\right),
%\end{multline}
\end{eqnarray}
\begin{eqnarray}
%\begin{multline}
[\Delta q\, \Delta \hat{K}\, D](x,z_h)= 
\Delta q(x) D(z_h)  +
\frac{\alpha_s}{2\pi} \int_x^1 \frac{dx'}{x'} \int_{z_h}^1 \frac{dz'}{z'} 
\Delta q \left(\frac{x}{x'}\right) \Delta C_{qq}(x',z') D\left(\frac{z_h}{z'}\right).
%\end{multline}
\end{eqnarray}

In Eqs. (\ref{apdif2}), (\ref{addif2}) 
$(1-1.5\omega_D)$, $\omega_D = 0.05\pm0.01$, is the factor accounting for
the  deuteron D-state contribution,
the quantity $R = \sigma_L/\sigma_T$ is taken from Ref. \cite{Rfactor},
$\Delta C_{qq}$ is the Wilson coefficient entering NLO QCD expression for
the polarized SIDIS structure function $g_1^h$ and can be found, for instance, 
in \cite{deFlorian_1997},
$C^2_{qq}=C^1_{qq}+C^L_{qq}$ is the Wilson coefficient entering the NLO QCD expression for
the unpolarized SIDIS structure function $F_2^h$, while the coefficients 
$C^1_{qq}$ and $ C^L_{qq}$ 
(entering $F_1^h$ and $F_L^h$) also can be found in \cite{deFlorian_1997}.
In our subsequent calculations we also use 
NLO parametrization GJR08 \cite{GJR} for unpolarized PDFs.

The equations allowing to find from the data on difference asymmetries 
the {\it n}-th moments
$
\Delta^{'}_n q\equiv \int_a^b dx\, x^{n-1} q(x)
$ 
of valence PDFs truncated to the accessible for measurement $x$ region $[a,b]$ 
look as
\begin{eqnarray}
\label{fmain2}
\Delta^{'}_n u_V \simeq \frac{1}{5}\frac{{\cal A}_p^{(n)}+{\cal A}_d^{(n)}}{L_{(n)1}-L_{(n)2}};
\quad  
\Delta^{'}_n d_V \simeq \frac{1}{5}\frac{4
{\cal A}_d^{(n)}-{\cal A}_p^{(n)}}{L_{(n)1}-L_{(n)2}},
\end{eqnarray}
where

\begin{eqnarray}
%\begin{multline}
\label{aint2}
{\cal A}_{p}^{(n)} &=& \sum_{i=1}^{N_{bins}} A_p^{\pi^+-\pi^-}
(\langle x_i\rangle){\Bigl |}_Z \int_{x_{i-1}}^{x_i}dx\, x^{n-1} 
 (1+R)^{-1} \int_Z^1 dz_h\, 
[(4u_V-d_V)\hat K (D_1-D_2)],\\
%\end{multline}
%\begin{multline}
{\cal A}_{d}^{(n)} &=&   
(1-1.5\,\omega_D )^{-1} \nonumber \\ &\times& \sum_{i=1}^{N_{bins}} A_p^{\pi^+-\pi^-}
(\langle x_i\rangle){\Bigl |}_Z  \int_{x_{i-1}}^{x_i}dx\, x^{n-1}  
 (1+R)^{-1} \int_Z^1 dz_h\, 
[(u_V+d_V)\hat K (D_1-D_2)].
%\end{multline}
\end{eqnarray}

The quantities $L_{(n)1}$, $L_{(n)2}$  are defined as 
\begin{eqnarray}
\label{lcoef2}
&&L_{(n)1}\equiv L_{(n)u}^{\pi^+}=L_{(n)\bar{u}}^{\pi^-}=L_{(n)\bar{d}}^{\pi^+}
=L_{(n)d}^{\pi^-},
\nonumber\\
&&L_{(n)2}\equiv L_{(n)d}^{\pi^+}=L_{{(n)}\bar{d}}^{\pi^-}=L_{(n)u}^{\pi^-}
=L_{(n)\bar{u}}^{\pi^+},\\
&&L_{(n)q}^h  
\equiv \int_Z^1 dz_h\left[D_q^h(z_h)+
\frac{\alpha_s}{2\pi}\int_{z_h}^1
\frac{dz'}{z'}\ \Delta_n C_{qq}(z')D_q^h(\frac{z_h}{z'})\right],\nonumber
\end{eqnarray}
where 
\begin{eqnarray}
\Delta_n C_{qq}(z)\equiv \int_0^1dx\ x^{n-1} \Delta C_{qq}(x,z).\nonumber
\end{eqnarray}

A lot of numerical tests performed in Refs. \cite{prd2004,prd2006} have shown 
that Eqs. (\ref{fmain2}) 
allow to reconstruct the truncated moments of polarized valence PDFs with a high 
precision even in the case of rather narrow HERMES $x$ region $0.023<x<0.6$, divided on
relatively small number of bins (nine bins only).

It is easy to see that for scenario with the broken $SU_f(2)$ symmetry 
one has instead of (\ref{apdif2}), (\ref{addif2}) the equations 
\begin{equation}
%\begin{multline}
\label{apdif_dss}
A_p^{\pi^+-\pi^-} = (1+R)  
\frac {(4\Delta u_V-\Delta d_V) \Delta \hat{K} (D_1-D_2) + 
\Delta d_V \Delta \hat{K} (D_1-\tilde{D_1})
}
{(4u_V-d_V) \hat{K} (D_1-D_2) +
d_V \hat{K} (D_1-\tilde{D_1})
},
%\end{multline}
\end{equation}
\begin{equation}
%\begin{multline}
\label{addif_dss}
A_d^{\pi^+-\pi^-} = (1+R)(1-\frac{3}{2}\omega_D)  
\frac
{(\Delta u_V+\Delta d_V) \Delta \hat{K} 
\bigl((D_1-D_2) + \frac{1}{3}(D_1-\tilde{D_1})\bigr)
}
{( u_V+d_V) \hat{K} 
\bigl((D_1-D_2) + \frac{1}{3}(D_1-\tilde{D_1})\bigr)
}.
%\end{multline}
\end{equation}
The generalization of Eqs. (\ref{fmain2})-(\ref{lcoef2}) on the case of 
broken $SU_f(2)$ symmetry (Eqs. (\ref{dssD1})-(\ref{dssD2})) is also
straightforward.

Both COMPASS \cite{compass2009,compass2010} and HERMES \cite{hermes} collaborations 
published the data only on asymmetries
$A_{p,d}^{\pi^{\pm}}$, while the published data on the pion difference asymmetries
$A_{p,d}^{\pi^{+}-\pi^{-}}$ are still\footnote{At present this work is in preparation
at COMPASS.} absent. That is why 
the special procedure was applied in \cite{prd2006}  
to construct asymmetries $A_{p,d}^{\pi^{+}-\pi^{-}}$ from the HERMES data on pion
production, and we repeat here this procedure for the COMPASS case.
Namely, in each {\it i}-th bin the pion difference asymmetries 
can be rewritten as 
\begin{eqnarray}
%\begin{multline}
\label{expansatz}
A^{\pi^+-\pi^-}(x_i) =\frac{R_i^{+/-}}{R_i^{+/-}-1}A^{\pi^+}(x_i)-
\frac{1}{R_i^{+/-}-1}A^{\pi^-}(x_i),
%\end{multline}
\end{eqnarray}
where $R^{+/-}_i$ is the ratio of unpolarized cross-sections  
for $\pi^{+}$ and  $\pi^{-}$ production: 
$
R^{+/-}_i = \sigma^{\pi^+}_{\text{unpol}}(x_i)/ \sigma^{\pi^-}_{\text{unpol}}(x_i)
=N^{\pi^+}_i/N^{\pi^-}_i.
$
As it was argued in Ref. \cite{prd2006} this relative quantity is very well reproduced by the  
the LEPTO generator of unpolarized events 
\cite{LEPTO}, which 
gives a good description of the fragmentation processes. So, we again use here the LEPTO
generator for this purpose.

Let us now discuss the question of $Q^2$ dependence of asymmetries and
its influence on the final results.
The point is that both DIS and SIDIS asymmetries very weakly depend on $Q^2$
(see, for instance,  Fig. 5 in Ref. \cite{compass-dis}), so that the approximation 
\begin{equation}
\label{approx-evol}
A(x_i,Q_i^2)\simeq A(x_i,Q_{0}^2)
\end{equation}
is commonly used (see, for example,  Refs. \cite{smc,hermes,compass2010}) for analysis of the 
DIS and SIDIS asymmetries. 
Nevertheless, for more comprehensive analysis,
it is useful to account for 
the corrections caused by 
the weak $Q^2$ dependence of the difference asymmetries,
i.e., to estimate the shifts
\begin{equation}
\label{corr}
\delta_i A_{p,d}^{\pi^+-\pi^-}=A_{p,d}^{\pi^+-\pi^-}(x_i,Q^2_{0})-
A_{p,d}^{\pi^+-\pi^-}(x_i,Q_i^2)
\end{equation}
in the difference asymmetries and their influence on the moments of valence PDFs.
To this end we first approximate 
r.h.s of Eq. (\ref{corr}) by the respective difference of ``theoretical'' asymmetries 
calculated 
with substitution of two novel parametrizations \cite{dssv,EPJ2010} on polarized PDFs 
(elaborated
with application of both DIS and SIDIS data) into the NLO QCD equations 
(\ref{apdif2}), (\ref{addif2}) 
(or (\ref{apdif_dss})-(\ref{addif_dss})),
and then average\footnote{Notice that the shifts in asymmetries as well as in the
final results on the moments of valence PDFs obtained with two applied parametrizations
differ very insignificantly from each other.} the obtained  
results on $\delta_i A_{p,d}^{\pi^+-\pi^-}$.
Adding the  calculated in this way $\delta_i A_{p,d}^{\pi^+-\pi^-}$ to the initial experimental
asymmetries $A_{p,d}^{\pi^+-\pi^-}(x_i,Q_i^2)$, we estimate the  
evolved from $Q_i^2$ to $Q^2_{0}$ asymmetries 
$A_{p,d}^{\pi^+-\pi^-}(x_i,Q_{0}^2)|_{evolved}$. 
Using the obtained in such a way\footnote
{Notice that the considered procedure of the asymmetry evolution is quite similar 
to the procedure used by SMC for the $\Gamma_{1p(d)}$ reconstruction 
(see Section  V$B$  in Ref. \cite{smc-overview}).
Notice also that the original procedure of $A_1$ evolution was elaborated 
in Refs. \cite{kotikov1997,kotikov1996}.
}
evolved asymmetries we extract the respective corrected moments of the valence PDFs 
$\Delta'_n q_V|_{corrected}$. 
Then we compare the 
corrected moments $\Delta'_n q_V|_{corrected}$  with the respective moments $\Delta'_n q_V$  
%from the previous  section (obtained without corrections due to evolution)
obtained  within the approximation (\ref{approx-evol}),
and calculate the respective shifts $\delta(\Delta'_n q_V)=
\Delta'_n q_V|_{corrected}-\Delta'_n q_V$ as well as the 
relative quantities
$\delta(\Delta'_n q_V)/\Delta'_n q_V$.

Of importance is the optimal choice of the scale $Q_0^2$ common for evolved asymmetries,   
allowing as much as possible to reduce the shifts in results due to evolution. 
Our experience
shows that for the combined analysis of COMPASS and HERMES data (see below) the optimal choice is
close to $Q_0^2=10\,\text{GeV}^2$.

\begin{figure}
\setlength{\tabcolsep}{1mm}
\caption{\footnotesize
Pion difference asymmetries $A_{p}^{\pi^+ - \pi^-}$ and $A_{d}^{\pi^+ - \pi^-}$ 
at $Q^2=10\,\text{GeV}^2$, 
constructed with Eq. (\ref{expansatz}) from the COMPASS data on 
$A_{p}^{\pi^{\pm}}$ and $A_{d}^{\pi^{\pm}}$ 
in the regions $0.004<x<0.7$ and $0.004<x<0.3$, respectively.
}
\begin{center}
\begin{tabular}{@{}cc@{}}
\includegraphics[height=5.5cm]{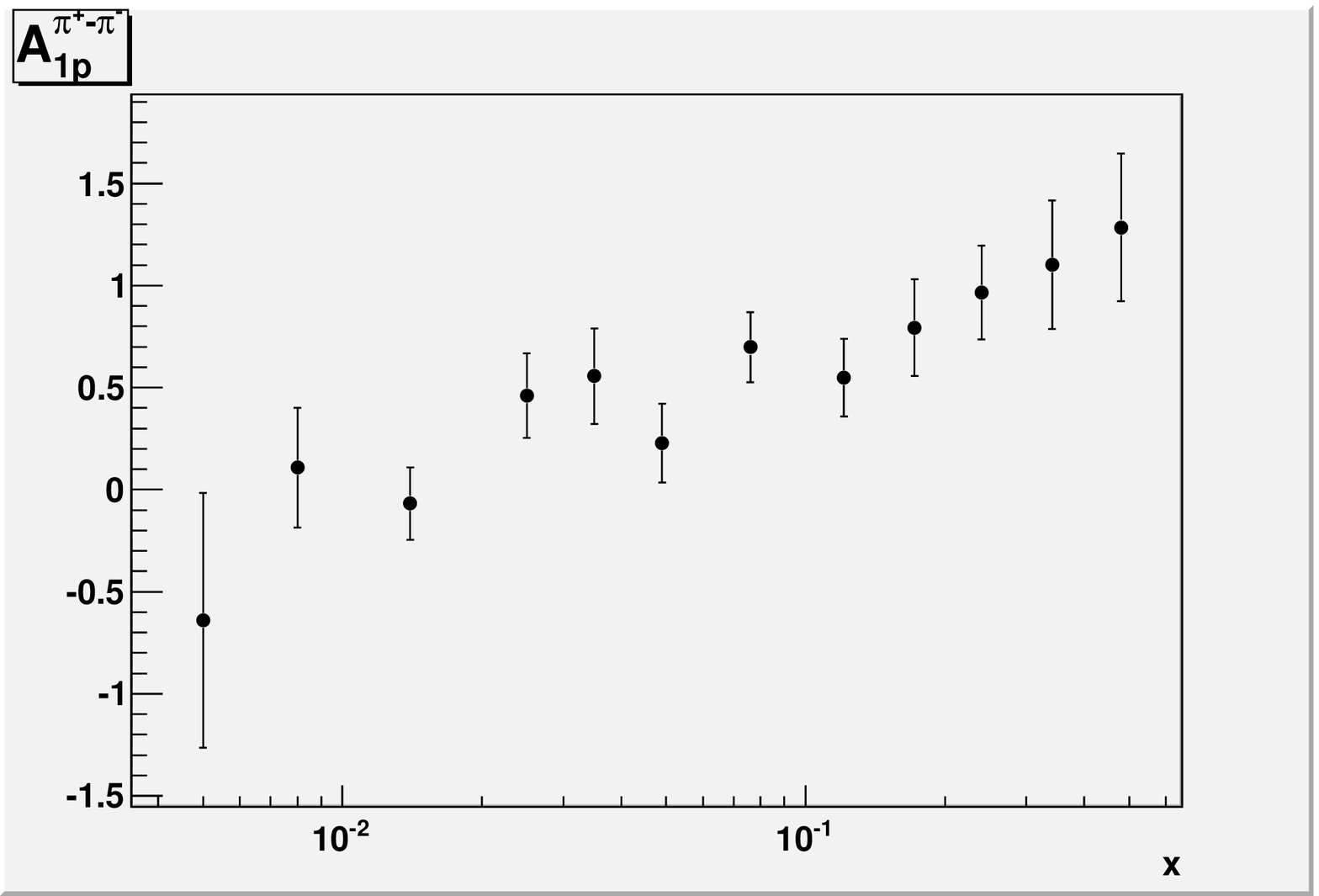} & 
\includegraphics[height=5.5cm]{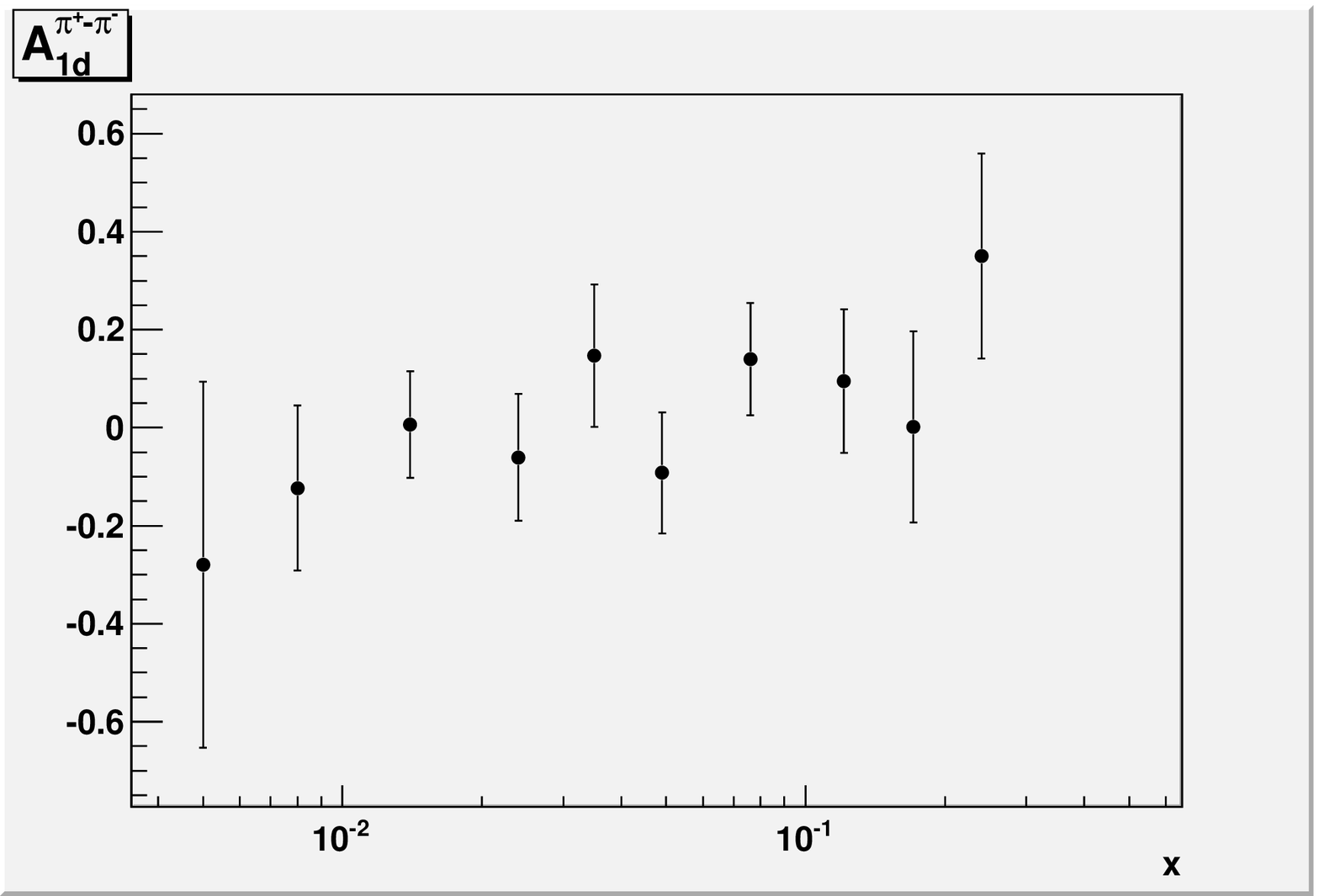}\\
\end{tabular}
\end{center}
\label{fig:asdif_compass}
\end{figure}

\begin{figure}
\setlength{\tabcolsep}{1mm}
\caption{\footnotesize
Pion difference asymmetries $A_{p}^{\pi^+ - \pi^-}$ and $A_{d}^{\pi^+ - \pi^-}$
at $Q^2=10\,\text{GeV}^2$, 
constructed with Eq. (\ref{expansatz}) from the HERMES data on $A_{p,d}^{\pi^{\pm}}$ 
in the region $0.023<x<0.6$.
}
\begin{center}
\begin{tabular}{@{}cc@{}}
\includegraphics[height=5.5cm]{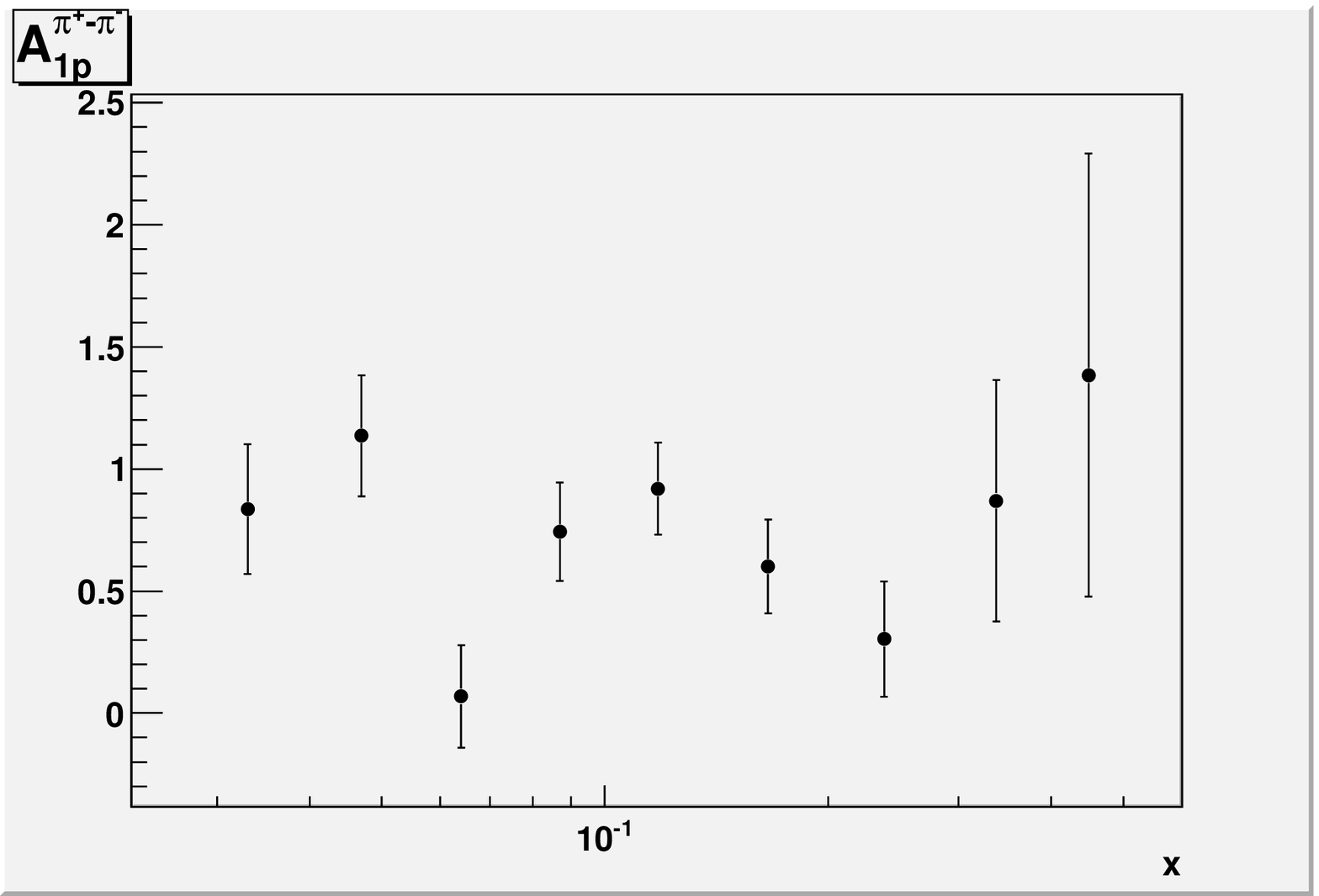} & 
\includegraphics[height=5.5cm]{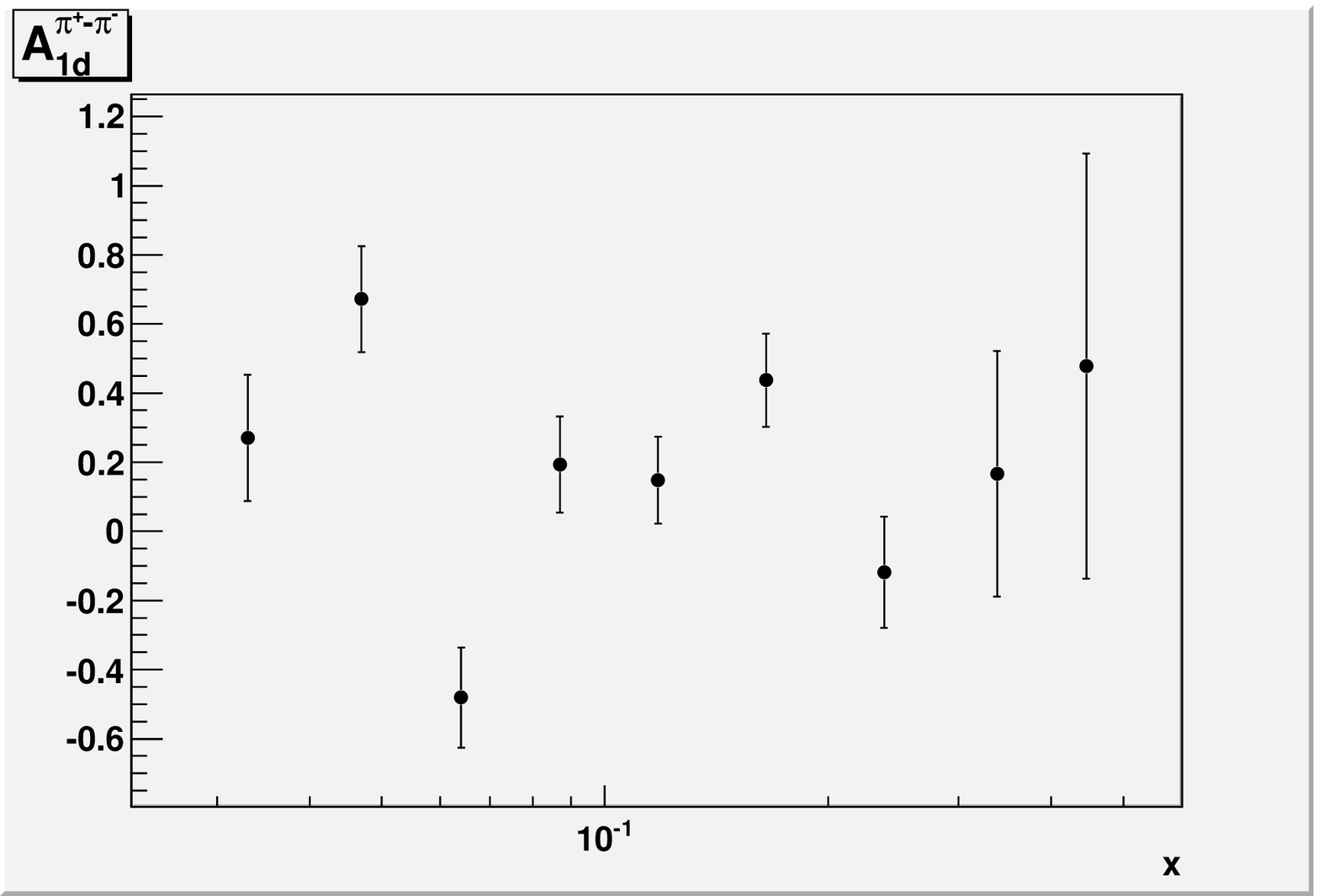}\\
\end{tabular}
\end{center}
\label{fig:asdif_hermes}
\end{figure}

The evolved to $Q^2=10\,GeV^2$ pion difference asymmetries constructed
with Eq. (\ref{expansatz}) from the
COMPASS \cite{compass2009,compass2010} and HERMES \cite{hermes} data on 
$A_{p,d}^{\pi^{\pm}}$ SIDIS asymmetries
are presented in Figs. \ref{fig:asdif_compass} and \ref{fig:asdif_hermes} respectively.

\begin{figure}
\setlength{\tabcolsep}{1mm}
\caption{\footnotesize
Pion difference asymmetries $A_{p}^{\pi^+ - \pi^-}$ and $A_{d}^{\pi^+ - \pi^-}$
at $Q^2=10\,\text{GeV}^2$, 
constructed with Eq. (\ref{expansatz}) from the COMPASS data on   
$A_{p,d}^{\pi^{\pm}}$ in the regions $0.004<x<0.7$, $0.004<x<0.3$ (for proton and deuteron
targets, respectively), and HERMES data
on $A_{p,d}^{\pi^{\pm}}$ in the region  $0.2<x<0.6$ (last three bins of HERMES).}
\begin{center}
\begin{tabular}{@{}cc@{}}
\includegraphics[height=5.5cm]{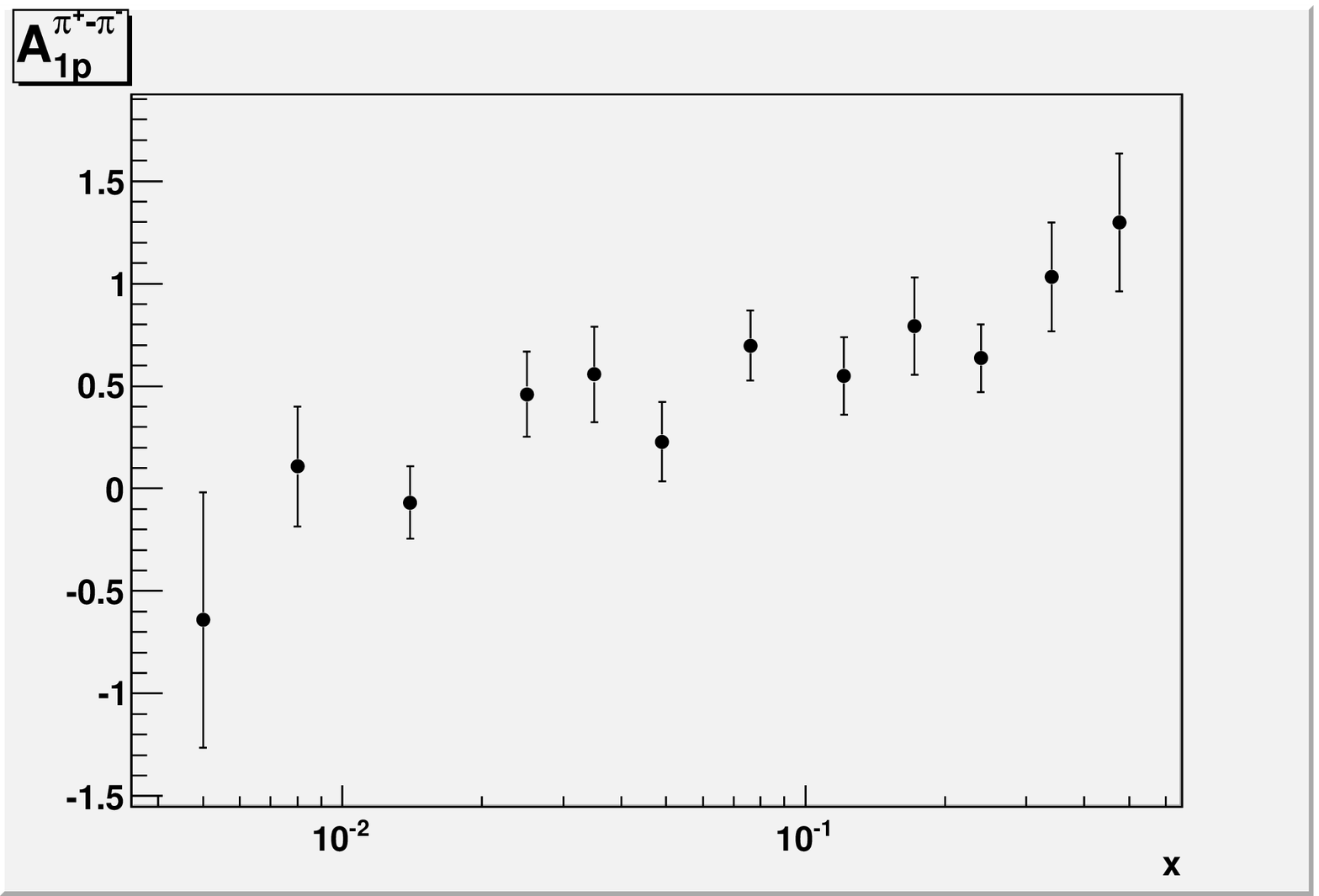} & 
\includegraphics[height=5.5cm]{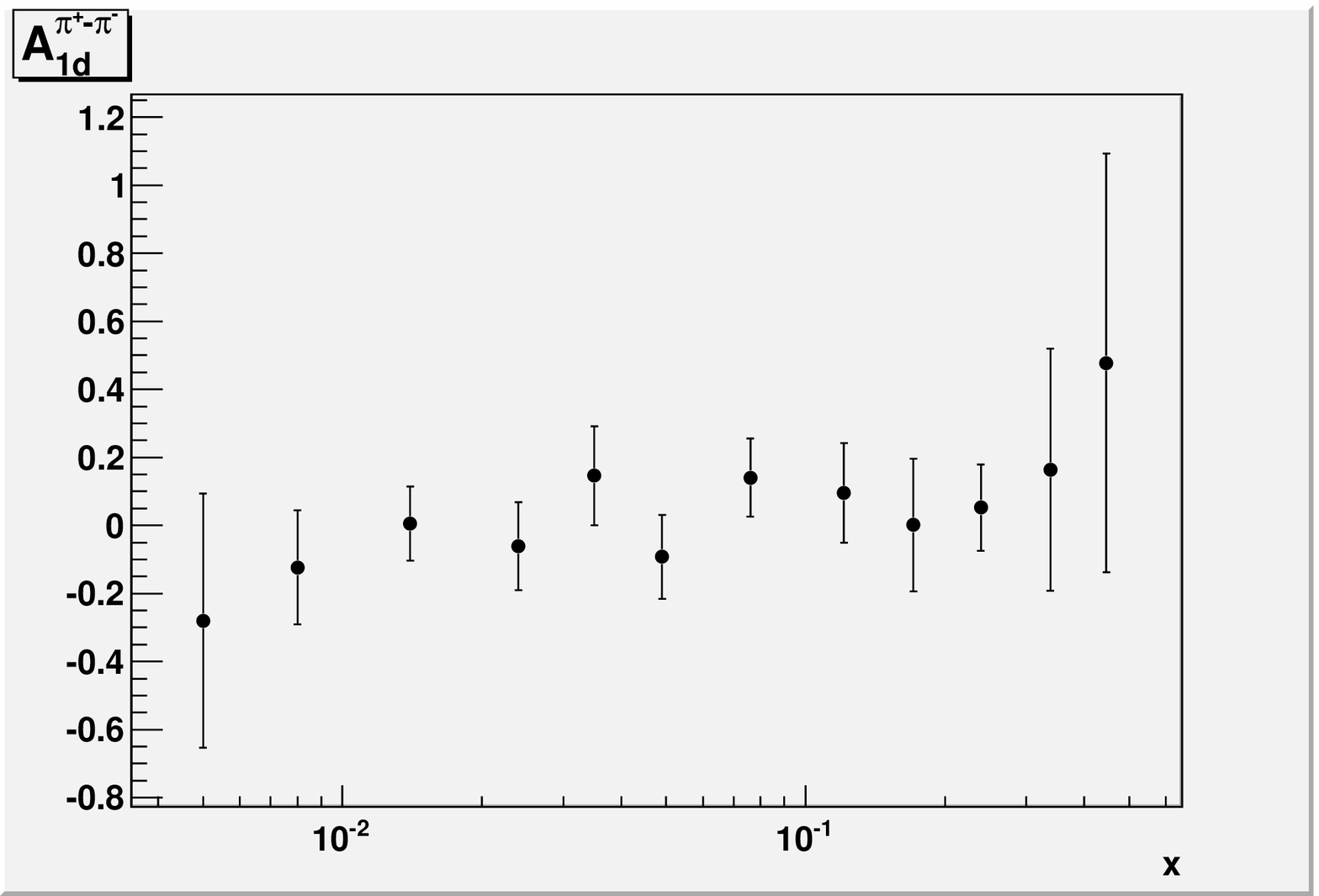}\\
\end{tabular}
\end{center}
\label{fig:asdif_cl3b}
\end{figure}

We perform the combined analysis of COMPASS \cite{compass2010,compass2009} and  
HERMES \cite{hermes} data on pion production with both proton and deuteron targets.
COMPASS collaboration published their data on $A_{p,d}^{\pi^{\pm}}$
in the Bjorken $x$ ranges $0.004<x<0.7$
and $0.004<x< 0.3$ for the proton and deuteron targets, respectively, while the HERMES 
data on these asymmetries were presented in the range $0.023<x<0.6$ for both
targets. The statistical summation of asymmetries $A_{p,d}^{\pi^{+} -\pi^{-}}$ 
(constructed with (\ref{expansatz})) is performed in accordance with 
the standard formulas 
\begin{equation}
%\begin{multline}
\label{stat_comb_value}
A_N^h|_{averaged} =
\frac{A_N^h|_{exp1}/(\delta A_N^h|_{exp1})^2 + A_N^h|_{exp2}/(\delta A_N^h|_{exp2})^2}
     {1/(\delta A_N^h|_{exp1})^2 + 1/(\delta A_N^h|_{exp2})^2},
%\end{multline}
\end{equation}
\begin{equation}
%\begin{multline}
\label{stat_comb_error}
(\delta A_N^h|_{averaged})^2 = \frac{1}{1/(\delta A_N^h|_{exp1})^2 
+ 1/(\delta A_N^h|_{exp2})^2}.
%\end{multline}
\end{equation}
At the same time one can apply Eqs. (\ref{stat_comb_value}), (\ref{stat_comb_error})
directly only for coinciding $x$ bins of different experiments.  
However, this is the case only for last three bins of COMPASS and HERMES experiments
we deal with (after proper extrapolation\footnote{Our experience show that 
%a such extrapolation 
this leads to negligible changes in the final results, 
irrespective of
the choice of the extrapolation procedure.
So, we apply here the simplest way of extrapolation, prescribing to $ A_{p,d}^{\pi^{\pm}} 
\bigl |_{HERMES}$ to be in the extended region $[0.4,0.7]$ the same as in the last HERMES 
$x$-bin 
 $[0.4,0.6]$.} of HERMES data in the last bin from $0.6$
to $0.7$ upper $x$ value). Besides, notice that 
for two last bins the COMPASS
published SIDIS data for deuteron target are still absent. That is why it is of especial
importance (and we do it first of all) to include in the analysis of COMPASS data the HERMES
data in the region $0.2<x<0.6$ (last three bins of HERMES). 
The results on the difference asymmetries obtained in such a way are presented 
in Fig. \ref{fig:asdif_cl3b}.

The respective results 
on the moments of polarized valence PDFs 
are presented in Table \ref{dqv_combined_last3bins}.
\begin{table}[h] \center
%combined last 3 bins
\caption{\footnotesize
Four first moments of polarized valence PDFs truncated to the region $0.004<x<0.7$
are presented at $Q^2=10\,GeV^2$.
The moments are obtained as a result of 
NLO QCD analysis of the combined data on $A_{p,d}^{\pi^+ -\pi^-}$
(see Fig. \ref{fig:asdif_cl3b}), 
constructed with Eq. (\ref{expansatz}) from the COMPASS data on   
$A_{p,d}^{\pi^{\pm}}$ in the regions $0.004<x<0.7$ (proton target),  $0.004<x<0.3$, 
(deuteron target), and HERMES data
on $A_{p,d}^{\pi^{\pm}}$ in the region  $0.2<x<0.6$ (last three bins of HERMES).
Capital letters A and B correspond to the application of AKK08 and DSS  
parametrizations for FFs, respectively. Rome numbers I and II correspond to  
the moments uncorrected and corrected due to evolution, respectively.
Besides, the relative corrections 
$\delta_{r}(\Delta'_n q_V)  \equiv \delta(\Delta'_n q_V)/\Delta'_n q_V$ 
caused by evolution are presented here.
}
\label{dqv_combined_last3bins}
\begin{tabular}{|c|c|c|c||c|c|c|}  \hline
\multicolumn{7}{|c|}{$\Delta'_n u_V $} \\ \hline
$n$ & $\text{A}_\text{I}$  & $\text{A}_\text{II}$ & $\delta_r(\Delta'_n u_V)$  
    & $\text{B}_\text{I}$  & $\text{B}_\text{II}$ & $\delta_r(\Delta'_n u_V)$ \\ \hline
$1$ & $ 0.731 \pm 0.087 $  & $ 0.695 \pm 0.087 $  &-3.8\% & $ 0.693 \pm 0.084 $ & $ 0.713 \pm 0.084 $ & 2.8\%  \\
$2$ & $ 0.166 \pm 0.024 $  & $ 0.167 \pm 0.024 $  & 0.8\% & $ 0.155 \pm 0.024 $ & $ 0.158 \pm 0.024 $ & 1.6\%  \\
$3$ & $ 0.055 \pm 0.010 $  & $ 0.055 \pm 0.010 $  & 1.3\% & $ 0.052 \pm 0.010 $ & $ 0.052 \pm 0.010 $ & 1.8\%  \\
$4$ & $ 0.022 \pm 0.005 $  & $ 0.022 \pm 0.005 $  & 1.5\% & $ 0.021 \pm 0.005 $ & $ 0.021 \pm 0.005 $ & 2.0\%  \\ \hline
\multicolumn{7}{|c|}{$\Delta'_n d_V $} \\ \hline
$n$ & $\text{A}_\text{I}$  & $\text{A}_\text{II}$ & $\delta_r(\Delta'_n d_V)$  
    & $\text{B}_\text{I}$  & $\text{B}_\text{II}$ & $\delta_r(\Delta'_n d_V)$ \\ \hline
$1$ & $ -0.519 \pm 0.162 $ & $ -0.524 \pm 0.162 $  & 0.9\%   & $ -0.473 \pm 0.157 $ & $ -0.481 \pm 0.157 $ & 1.7\%  \\
$2$ & $ -0.100 \pm 0.054 $ & $ -0.102 \pm 0.054 $  & 1.8\%   & $ -0.090 \pm 0.051 $ & $ -0.092 \pm 0.051 $ & 2.7\%  \\
$3$ & $ -0.029 \pm 0.023 $ & $ -0.030 \pm 0.023 $  & 2.5\%   & $ -0.026 \pm 0.022 $ & $ -0.027 \pm 0.022 $ & 3.7\%  \\
$4$ & $ -0.011 \pm 0.011 $ & $ -0.011 \pm 0.011 $  & 3.1\%   & $ -0.010 \pm 0.010 $ & $ -0.010 \pm 0.010 $ & 4.4\%  \\  \hline
\end{tabular}
\end{table}

\begin{figure}
\setlength{\tabcolsep}{1mm}
\caption{\footnotesize
Pion difference asymmetries $A_{p}^{\pi^+ - \pi^-}$ and $A_{d}^{\pi^+ - \pi^-}$
at $Q^2=10\,\text{GeV}$,
constructed with Eq. (\ref{expansatz}) from the COMPASS 
and HERMES data on $A_{p,d}^{\pi^{\pm}}$ in the entire $x$-regions accessible for measurement
($0.004<x<0.7$, $0.004<x<0.3$ for COMPASS and $0.023<x<0.6$ for HERMES).}
\begin{center}
\begin{tabular}{@{}cc@{}}
\includegraphics[height=5.5cm]{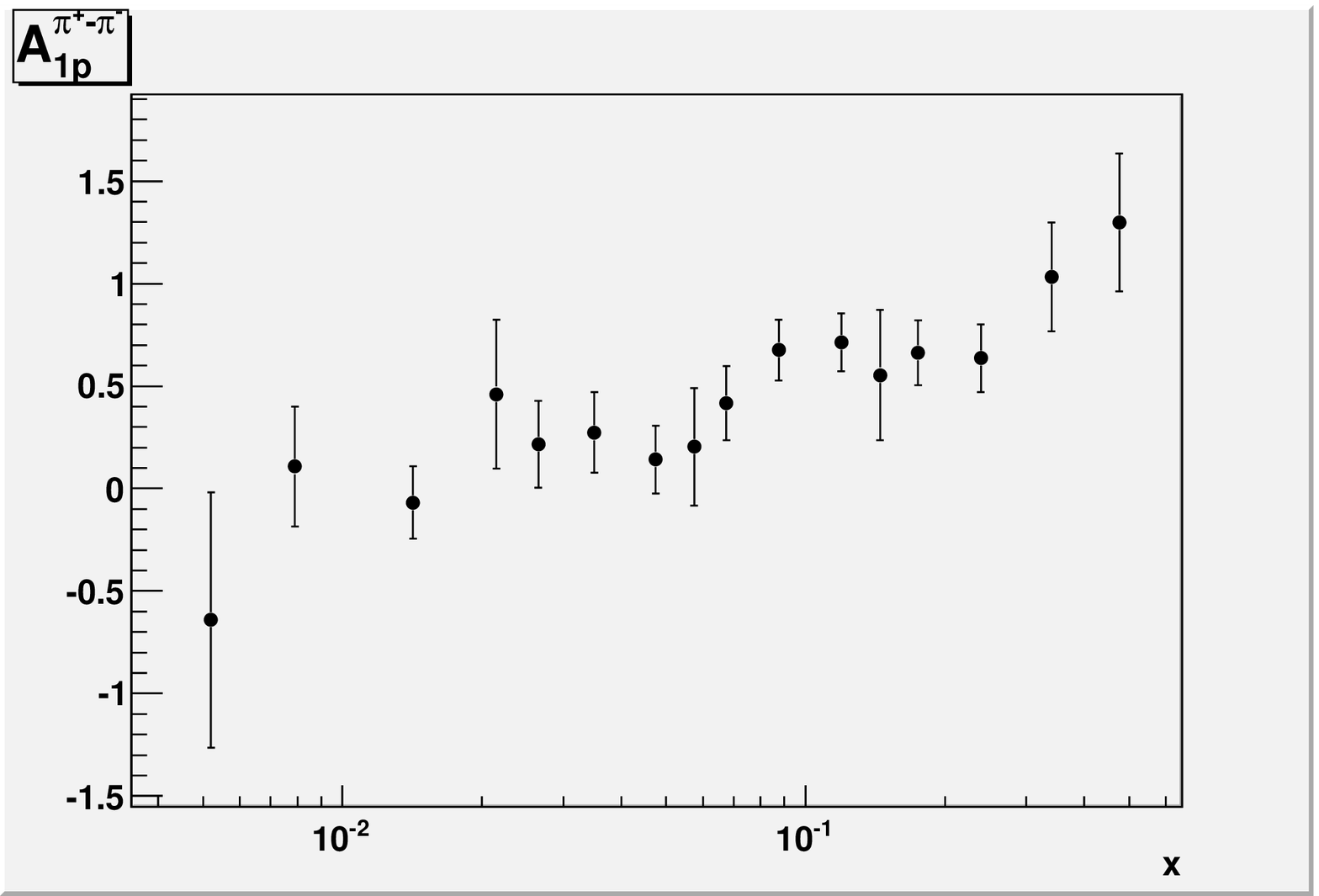} & 
\includegraphics[height=5.5cm]{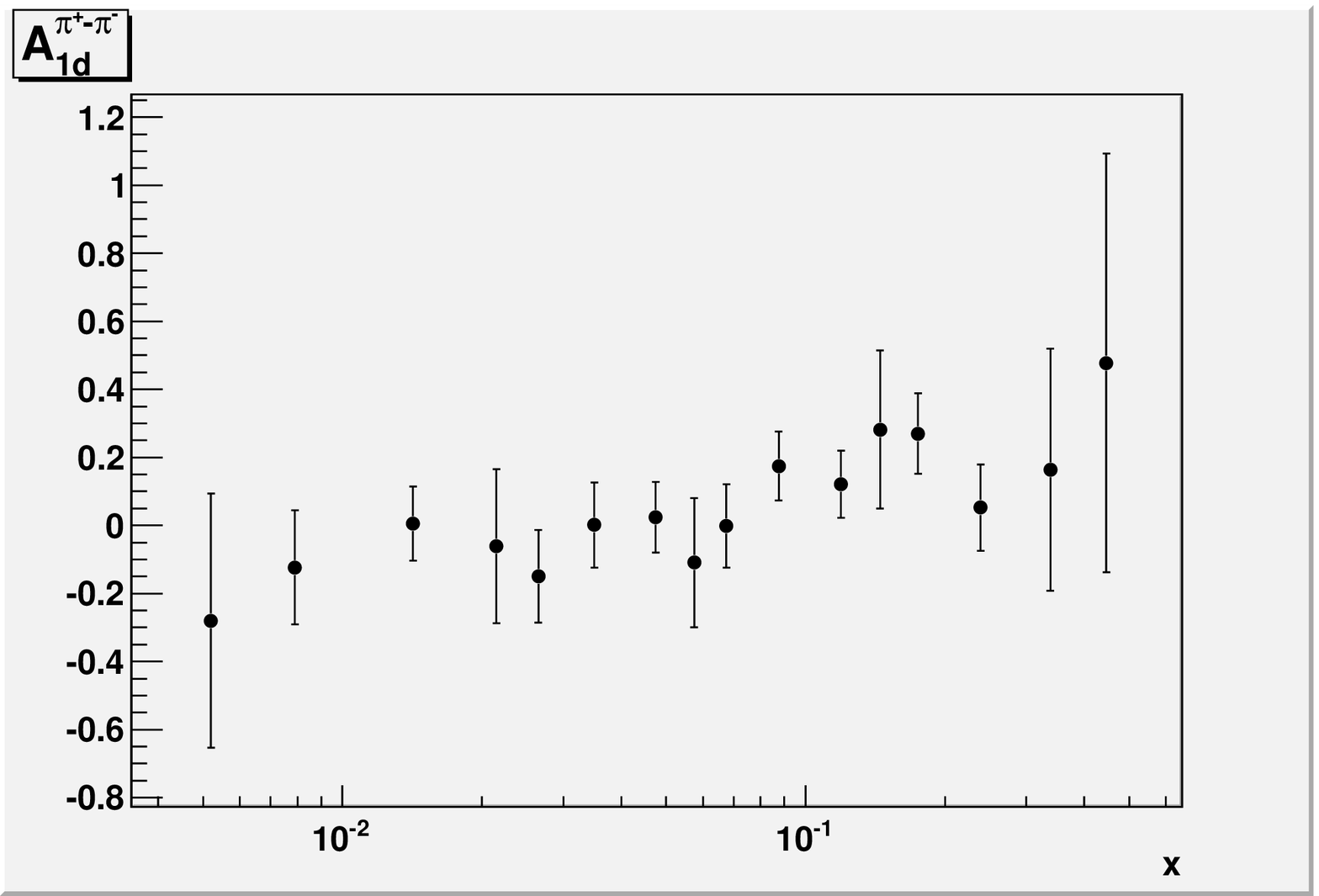}\\
\end{tabular}
\end{center}
\label{fig:asdif_cfull}
\end{figure}

On the other hand, to maximally increase the available statistics it is, certainly, 
very desirable to perform the complete combined analysis, using the 
COMPASS and HERMES data taken in the entire $x$ regions accessible for these experiments. 
To this end we have elaborated the special procedure allowing to combine
the data on asymmetries coming from experiments with different binnings -- see the Appendix.
The results on the difference asymmetries obtained in such a way are presented 
in Fig. \ref{fig:asdif_cfull}.

The respective results on the moments of polarized valence PDFs are listed
in Table \ref{dqv_combined_full}.   
\begin{table}[h] \center
%combined full
\caption{\footnotesize
Four first moments of polarized valence PDFs truncated to the region $0.004<x<0.7$
are presented at $Q^2=10\,\text{GeV}^2$.
The moments are obtained as a result of 
NLO QCD analysis of the combined data on $A_{p,d}^{\pi^+ -\pi^-}$
(see Fig. \ref{fig:asdif_cfull}), 
constructed with Eq. (\ref{expansatz}) from the COMPASS 
and HERMES data on $A_{p,d}^{\pi^{\pm}}$ in the entire $x$-regions accessible for measurement
($0.004<x<0.7$, $0.004<x<0.3$ for COMPASS and $0.023<x<0.6$ for HERMES).
Capital letters A and B correspond to the application of AKK08 and DSS  
parametrizations for FFs, respectively. Rome numbers I and II correspond to  
the moments uncorrected and corrected due to evolution, respectively.
Besides, the relative corrections 
$\delta_{r}(\Delta'_n q_V)  \equiv \delta(\Delta'_n q_V)/\Delta'_n q_V$ 
caused by evolution are presented here.
}
\label{dqv_combined_full}
\begin{tabular}{|c|c|c|c||c|c|c|}  \hline
\multicolumn{7}{|c|}{$\Delta'_n u_V $} \\ \hline
$n$ & $\text{A}_\text{I}$  & $\text{A}_\text{II}$ & $\delta_r(\Delta'_n u_V)$  
    & $\text{B}_\text{I}$  & $\text{B}_\text{II}$ & $\delta_r(\Delta'_n u_V)$ \\ \hline
$1$ & $ 0.712 \pm 0.078 $  & $ 0.660 \pm 0.078 $  &-4.3\% & $ 0.683 \pm 0.076 $ & $ 0.711 \pm 0.076 $ & 4.0\%  \\
$2$ & $ 0.166 \pm 0.023 $  & $ 0.168 \pm 0.023 $  & 0.8\% & $ 0.156 \pm 0.024 $ & $ 0.159 \pm 0.024 $ & 1.7\%  \\
$3$ & $ 0.055 \pm 0.010 $  & $ 0.056 \pm 0.010 $  & 1.4\% & $ 0.052 \pm 0.010 $ & $ 0.053 \pm 0.010 $ & 1.8\%  \\
$4$ & $ 0.022 \pm 0.005 $  & $ 0.022 \pm 0.005 $  & 1.6\% & $ 0.021 \pm 0.005 $ & $ 0.021 \pm 0.005 $ & 2.0\%  \\ \hline
\multicolumn{7}{|c|}{$\Delta'_n d_V $} \\ \hline
$n$ & $\text{A}_\text{I}$  & $\text{A}_\text{II}$ & $\delta_r(\Delta'_n d_V)$  
    & $\text{B}_\text{I}$  & $\text{B}_\text{II}$ & $\delta_r(\Delta'_n d_V)$ \\ \hline
$1$ & $ -0.414 \pm 0.149 $ & $ -0.427 \pm 0.149 $  & 3.0\%   & $ -0.376 \pm 0.145 $ & $ -0.381 \pm 0.145 $ & 1.4\%  \\
$2$ & $ -0.087 \pm 0.053 $ & $ -0.089 \pm 0.053 $  & 2.5\%   & $ -0.078 \pm 0.051 $ & $ -0.080 \pm 0.051 $ & 2.7\%  \\
$3$ & $ -0.027 \pm 0.023 $ & $ -0.028 \pm 0.023 $  & 2.8\%   & $ -0.024 \pm 0.022 $ & $ -0.025 \pm 0.022 $ & 3.8\%  \\
$4$ & $ -0.010 \pm 0.011 $ & $ -0.011 \pm 0.011 $  & 3.3\%   & $ -0.009 \pm 0.010 $ & $ -0.010 \pm 0.010 $ & 4.6\%  \\  \hline
\end{tabular}
\end{table}

In Tables \ref{dqv_combined_last3bins}, \ref{dqv_combined_full} both scenarios 
Eqs. (\ref{D_fav}), (\ref{D_unfav}) (AKK08 parametrization) 
and Eqs. 
%(\ref{dssD1}), (\ref{dssD11}), (\ref{dssD2}) 
(\ref{dssD1})-(\ref{dssD2}) 
(DSS parametrization) 
for FFs are considered. One can see that the results are consistent within the errors.
Besides, for each scenario we present the results obtained with and without 
corrections due to evolution of asymmetries. It is seen that the difference is not
too significant 
(the relative corrections $\delta(\Delta'_n q_V)/\Delta'_n q_V$ 
take the small values).

Thus, we estimated in NLO QCD the 
contributions of valence quarks (first moments of polarized valence PDFs) 
to the  nucleon spin.
Let us now estimate the respective 
contributions of light sea quarks.
We do it in two different ways.

Within the first procedure one first of all uses some NLO QCD
parametrization on the polarized PDFs 
to estimate the quantities $\Delta'_1 q + \Delta'_1 \bar{q}\,\,(q=u,d)$.
Since the sums $\Delta q(x) + \Delta \bar q(x) \,\,(q=u,d)$ are well fitted by the precise 
purely
inclusive DIS data (these quantities are considered as relatively well known and practically 
are the same for the different modern parametrizations) it is not especially
important which   
parametrization one applies for this purpose (here we use the most popular and widely
cited DSSV \cite{dssv} parametrization). 
Then, having both $(\Delta_1' q + \Delta_1' \bar{q})|_{\text{parametrization}}
\,\,(q=u,d)$
and (see Tables \ref{dqv_combined_last3bins} and \ref{dqv_combined_full}) $\Delta'_1 q_V
 \,\,(q=u,d)$  quantities     
one easily gets the truncated first moments of sea $u$ and $d$ quarks, applying
the obvious relation
\begin{equation}
\label{obvious}
\Delta_1' \bar{q} = \frac{1}{2} [(\Delta_1' q + \Delta_1' 
\bar{q}) - \Delta'_1 q_V]. 
\end{equation}
The first moments $\Delta_1' \bar{u}$, $\Delta_1' \bar{d}$ obtained in this way,
as well as their differences and sums are presented in Tables
\ref{sea_cl3b_dssv} and \ref{sea_cfull_dssv}.
\begin{table}[h] \center
%sea, combined last 3 bins, DSSV
\caption{\footnotesize
First moments of polarized sea PDFs truncated to the region $0.004<x<0.7$
are presented at $Q^2=10\,GeV^2$,
as well as their sums and differences.
The moments are obtained with application of Eq. (\ref{obvious}), 
where DSSV parametrization is used to estimate 
$(\Delta_1' q + \Delta_1' \bar{q})|_{\text{parametrization}}$, while
the first moments of valence PDFs are taken from Table \ref{dqv_combined_last3bins} 
(HERMES data only from three last bins are applied).
Capital letters A and B correspond to the application of AKK08 and DSS  
parametrizations for FFs, respectively. Rome numbers I and II correspond to  
the moments uncorrected and corrected due to evolution, respectively.
}
\label{sea_cl3b_dssv}
\begin{tabular}{|c|r|r|r|r|} \hline
 &\multicolumn{1}{c|}{$\text{A}_\text{I}$} &\multicolumn{1}{c|}{$\text{A}_\text{II}$} 
 &\multicolumn{1}{c|}{$\text{B}_\text{I}$} &\multicolumn{1}{c|}{$\text{B}_\text{II}$}      \\ \hline
$\Delta'_1 \bar u$ 
& $ 0.018 \pm 0.044 $  & $ 0.036 \pm 0.044 $  & $ 0.037 \pm 0.042 $  & $ 0.027 \pm 0.042 $ \\  
$\Delta'_1 \bar d$ 
& $ 0.065 \pm 0.081 $  & $ 0.067 \pm 0.081 $  & $ 0.042 \pm 0.079 $  & $ 0.046 \pm 0.079 $ \\
$\Delta'_1 \bar u + \Delta_1 \bar d$ 
& $ 0.082 \pm 0.092 $  & $ 0.102 \pm 0.092 $  & $ 0.078 \pm 0.089 $  & $ 0.072 \pm 0.089 $ \\
$\Delta'_1 \bar u - \Delta_1 \bar d$ 
& $-0.047 \pm 0.092 $  & $-0.032 \pm 0.092 $  & $-0.005 \pm 0.089 $  & $-0.019 \pm 0.089 $ \\ \hline
\end{tabular}
\end{table}
\begin{table}[h] \center
%sea, combined full, DSSV
\caption{\footnotesize
First moments of polarized sea PDFs truncated to the region $0.004<x<0.7$
are presented at $Q^2=10\,GeV^2$,
as well as their sums and differences.
The moments are obtained with application of Eq. (\ref{obvious}), 
where DSSV parametrization is used to estimate 
$(\Delta_1' q + \Delta_1' \bar{q})|_{\text{parametrization}}$, while
the first moments of valence PDFs are taken from Table \ref{dqv_combined_full}
(all data on pion production of COMPASS and HERMES are combined).
Capital letters A and B correspond to the application of AKK08 and DSS  
parametrizations for FFs, respectively. Rome numbers I and II correspond to  
the moments uncorrected and corrected due to evolution, respectively.
}
\label{sea_cfull_dssv}
\begin{tabular}{|c|r|r|r|r|} \hline
 &\multicolumn{1}{c|}{$\text{A}_\text{I}$} &\multicolumn{1}{c|}{$\text{A}_\text{II}$} 
 &\multicolumn{1}{c|}{$\text{B}_\text{I}$} &\multicolumn{1}{c|}{$\text{B}_\text{II}$}      \\ \hline
$\Delta'_1 \bar u$ 
& $ 0.027 \pm 0.039 $  & $ 0.053 \pm 0.039 $  & $ 0.042 \pm 0.038 $  & $ 0.028 \pm 0.038 $ \\  
$\Delta'_1 \bar d$ 
& $ 0.012 \pm 0.075 $  & $ 0.019 \pm 0.075 $  & $-0.007 \pm 0.073 $  & $-0.004 \pm 0.073 $ \\
$\Delta'_1 \bar u + \Delta'_1 \bar d$ 
& $ 0.039 \pm 0.084 $  & $ 0.072 \pm 0.084 $  & $ 0.035 \pm 0.082 $  & $ 0.023 \pm 0.082 $ \\
$\Delta'_1 \bar u - \Delta'_1 \bar d$ 
& $ 0.015 \pm 0.084 $  & $ 0.034 \pm 0.084 $  & $ 0.048 \pm 0.082 $  & $ 0.032 \pm 0.082 $ \\ \hline
\end{tabular}
\end{table}

The idea of alternative  procedure for investigation of the sea contributions to the 
nucleon spin in NLO QCD is based on the proper application of $SU_f(2)$ (Bjorken sum rule)
and $SU_f(3)$ sum rules:  
\be
%\begin{multline}
\label{bjorken_sum_rule}
a_3 &\equiv& (\Delta_1 u + \Delta_1 \bar{u}) - (\Delta_1 d + \Delta_1 \bar{d})  
= \left|\frac{g_A}{g_V}\right|  = F+D = 1.2670 \pm 0.0035, \\
%\end{multline}
%\end{equation}
%\begin{multline}
%\begin{equation}
\label{su3_sum_rule}
a_8 &\equiv& \Delta_1 u + \Delta_1 \bar{u} + \Delta_1 d + 
\Delta_1 \bar{d} - 2(\Delta_1 s + \Delta_1 \bar{s})   = 3F-D = 0.585 \pm 0.025.
%\end{multline}
\ee

Bjorken sum rule (\ref{bjorken_sum_rule}) 
rewritten in terms of
valence and sea distributions produces quite good approximation \cite{prd2004,prd2003} 
(see Ref. \cite{our-review} for review) 
\begin{equation}
\label{bsr_approx}
{
\Delta_1\bar{u}-\Delta_1\bar{d} \simeq \frac{1}{2}\ \left |\frac{g_A}{g_V}\right |
-\frac{1}{2}\ (\Delta_1' u_V-\Delta_1' d_V)
}
\end{equation}
for the difference of {\it full} (not truncated!) moments
$\Delta_1\bar{u}$ and $\Delta_1\bar{d}$
even in the case of rather narrow HERMES $x$-range, while for the wide COMPASS  
$x$-range we deal with here this approximation works very well --
see the respective numerical tests in Ref. \cite{prd2004}.
The point is that since 
the valence PDFs (contrary to the sea PDFs)
%gather far from the low boundary $x=0$,
are suppressed near the low boundary $x=0$ (so that $\Delta q =\Delta q_V+\Delta \bar q
\to \Delta \bar q$ as $x\to 0$),
the omitted in r.h.s. of Eq. (\ref{bsr_approx}) term $\int_0^a dx\,(\Delta u_V - \Delta d_V)$ 
is small even for HERMES low $x$ boundary $a=0.023$, and becomes  
really negligible for COMPASS $a=0.004$ we deal with here. 
In turn, another omitted term $\int_{0.7}^1 dx\,(\Delta u_V - \Delta d_V)$ in r.h.s of (\ref{bsr_approx}) 
is also negligible since all PDFs 
just die out at so high $x$ values.

On the other hand, to estimate the sum of {\it full} moments in NLO QCD
we use the sum rule (\ref{su3_sum_rule}) and purely inclusive DIS data  
on the first moment $\Gamma_{1}^d$ of deuteron structure function $g_{1d}$ (measured
with high precision).
To this end we use the NLO QCD expression for $\Gamma_1^N \equiv (1-1.5\,\omega_D)^{-1} 
\Gamma_{1}^d$:
\begin{equation}
%\begin{multline}
\Gamma_1^N \equiv (1-1.5\,\omega_D)^{-1} \Gamma_{1}^d =\frac{1}{2}(\Gamma_1^p + \Gamma_1^n) 
=
\left(1-\frac{\alpha_s(Q^2)}{\pi}\right)\left(\frac{1}{36}a_8 + 
\frac{1}{9}\Delta_1\Sigma(Q^2)\right),
%\end{multline}
\end{equation}
which produces very good approximation for $\Delta_1\bar{u}+\Delta_1\bar{d}$:
\begin{equation}
%\begin{multline}
\label{sea_sum}
\Delta_1 \bar{u} + \Delta_1 \bar{d} \simeq \left( 3 \left(1+\frac{\alpha_s}{\pi}\right) 
\Gamma_1^N + 
\frac{1}{12}a_8 \right) 
- \frac{1}{2}(\Delta_1' u_V + \Delta_1' d_V),
%\end{multline}
\end{equation}
where we again omitted the small contributions of valence PDFs 
$\int_0^{0.004} dx\, (\Delta u_V + \Delta d_V)$ and 
$\int_{0.7}^1 dx\,(\Delta u_V + \Delta d_V)$). 
We use in Eq. (\ref{sea_sum}) the numerical value of $\Gamma_1^N$ 
taken from the COMPASS paper \cite{compass2007}:
$$ \Gamma_1^N = 0.051 \pm 0.003 \pm 0.006. $$

The obtained with Eqs. (\ref{bsr_approx}), (\ref{sea_sum}) results on the sums and 
differences of 
the first moments of sea PDFs, 
as well as on the moments themselves in separation
are presented in Tables \ref{sea_cl3b_sumrules} and \ref{sea_cfull_sumrules}.
\begin{table}[h] \center
%sea, combined last 3 bins, BSR+...
\caption{\footnotesize
Sums and differences of the first moments of polarized sea PDFs, 
as well as the moments themselves,
obtained in NLO QCD at $Q^2=10\,GeV^2$ within the approximations 
(\ref{bsr_approx}) and (\ref{sea_sum}).
The truncated first moments of valence PDFs are taken from the 
Table \ref{dqv_combined_last3bins} 
(HERMES data only from three last bins are applied).
Capital letters A and B correspond to the application of AKK08 and DSS  
parametrizations for FFs, respectively. Rome numbers I and II correspond to  
the moments uncorrected and corrected due to evolution, respectively.
}
\label{sea_cl3b_sumrules}
\begin{tabular}{|c|r|r|r|r|} \hline
 &\multicolumn{1}{c|}{$\text{A}_\text{I}$} &\multicolumn{1}{c|}{$\text{A}_\text{II}$} 
 &\multicolumn{1}{c|}{$\text{B}_\text{I}$} &\multicolumn{1}{c|}{$\text{B}_\text{II}$}      \\ \hline
$\Delta_1 \bar u$ 
& $ 0.059 \pm 0.045 $  & $ 0.077 \pm 0.045 $  & $ 0.078 \pm 0.043 $  & $ 0.068 \pm 0.043 $ \\
$\Delta_1 \bar d$ 
& $ 0.050 \pm 0.082 $  & $ 0.053 \pm 0.082 $  & $ 0.027 \pm 0.079 $  & $ 0.031 \pm 0.079 $ \\
$\Delta_1 \bar u + \Delta_1 \bar d$ 
& $ 0.109 \pm 0.095 $  & $ 0.129 \pm 0.095 $  & $ 0.105 \pm 0.092 $  & $ 0.099 \pm 0.092 $ \\
$\Delta_1 \bar u - \Delta_1 \bar d$ 
& $ 0.009 \pm 0.092 $  & $ 0.024 \pm 0.092 $  & $ 0.051 \pm 0.089 $  & $ 0.037 \pm 0.089 $ \\
\hline
\end{tabular}
\end{table}
\begin{table}[h] \center
%sea, combined full, BSR+...
\caption{\footnotesize
Sums and differences of the first moments of polarized sea PDFs, 
as well as the moments themselves,
obtained in NLO QCD at $Q^2=10\,GeV^2$ within the approximations 
(\ref{bsr_approx}) and (\ref{sea_sum}).
The truncated first moments of valence PDFs are taken from the Table 
\ref{dqv_combined_full}
(all data on pion production of COMPASS and HERMES are combined).
Capital letters A and B correspond to the application of AKK08 and DSS  
parametrizations for FFs, respectively. Rome numbers I and II correspond to  
the moments uncorrected and corrected due to evolution, respectively.
}
\label{sea_cfull_sumrules}
\begin{tabular}{|c|r|r|r|r|} \hline
 &\multicolumn{1}{c|}{$\text{A}_\text{I}$} &\multicolumn{1}{c|}{$\text{A}_\text{II}$} 
 &\multicolumn{1}{c|}{$\text{B}_\text{I}$} &\multicolumn{1}{c|}{$\text{B}_\text{II}$}      \\ \hline
$\Delta_1 \bar u$ 
& $ 0.068 \pm 0.041 $  & $ 0.094 \pm 0.041 $  & $ 0.083 \pm 0.040 $  & $ 0.069 \pm 0.040 $ \\ 
$\Delta_1 \bar d$ 
& $-0.002 \pm 0.075 $  & $ 0.004 \pm 0.075 $  & $-0.021 \pm 0.073 $  & $-0.019 \pm 0.073 $ \\
$\Delta_1 \bar u + \Delta_1 \bar d$ 
& $ 0.066 \pm 0.087 $  & $ 0.099 \pm 0.087 $  & $ 0.061 \pm 0.085 $  & $ 0.050 \pm 0.085 $ \\ 
$\Delta_1 \bar u - \Delta_1 \bar d$ 
& $ 0.071 \pm 0.084 $  & $ 0.090 \pm 0.084 $  & $ 0.104 \pm 0.082 $  & $ 0.087 \pm 0.082 $ \\ 
\hline
\end{tabular}
\end{table}

Looking at Tables \ref{sea_cl3b_dssv}, \ref{sea_cfull_dssv}, and 
Tables\footnote{While the values of $\Delta_1 \bar d$ listed in 
in Tables \ref{sea_cl3b_sumrules}, \ref{sea_cfull_sumrules} 
are just zeros within the errors, the values of $\Delta_1 \bar u$ 
are compatible with zero within $2 \sigma$.}
\ref{sea_cl3b_sumrules}, \ref{sea_cfull_sumrules} 
one can draw an unexpected conclusion, that irrespective of the 
procedure used in 
%considered here way of 
the SIDIS data analysis 
the first moments of sea PDFs are consistent with zero  
\footnote{Certainly, because of 
evolution these quantities
still can deviate from zero 
at values of $Q^2$ distinct from 
considered here $Q^2=10\,\text{GeV}^2$. However, in the wide range of  $Q^2$
really available to experiment they are still negligible.
} 
within the errors. 
In particular, in contrast with the different model predictions 
\cite{bhalerao,peng,bsb,waka,waka_1,kuma}
the polarized sea asymmetry $\Delta_1 {\bar u} - \Delta_1 {\bar d}$
appear to be just zero within the errors. 
It is of importance, because some of these models (see Refs.~\cite{peng,waka} 
and references therein)
predict that this asymmetry is even larger in absolute value 
than the unpolarized sea asymmetry ${\bar u}-{\bar d}$ (which takes quite considerable
value allowing to explain a large violation of the 
Gottfried sum rule).

{\it In conclusion}, let us briefly discuss the obtained results.

The first moments of the polarized valence PDFs 
truncated to the wide Bjorken $x$ region $0.004<x<0.7$
are directly (without any fitting procedure) extracted in NLO QCD from both 
COMPASS and HERMES polarized SIDIS data. 
To this end we apply two scenarios for the fragmentation functions and
two ways to combine the COMPASS and HERMES data on pion production (one 
of them is based on the proposed special procedure, allowing to combine
all data coming from experiments with different binnings).
In turn, the obtained results on the valence PDFs 
have allowed us to estimate  in two ways
the contributions of light sea quarks 
to the proton spin which, surprisingly, occur compatible with zero\footnote{
Notice that similar conclusion was made in the recent COMPASS paper \cite{compass2010},
where the truncated moment of sea $u$ quark is zero within the errors, while
the moment of $d$ quark very slightly differs from zero. However, first, the analysis
in \cite{compass2010} was performed only in LO QCD. 
Second, the moments studied there
were truncated to the rather narrow region $0.004<x<0.3$ (because of lack of the
deuteron data in the last two bins). 
Third, all set of FFs has been used, 
%(in particular poorly known FFs 
%$D_q^{K^\pm}$ and $D_g^h$), 
while we apply only well known pion FFs.
Fourth, the corrections due to evolution were not taken into account in \cite{compass2010}  
(approximation (\ref{approx-evol}) was applied).
} within the errors. 
Certainly, this conclusion should be considered as still preliminary,
since the results on the sea contributions are obtained for the restricted Bjorken $x$ region. 
Nevertheless, 
its degree of reliability is high enough due to the discussed above advantage
of approximations (\ref{bsr_approx}), (\ref{sea_sum}) to the full moments, 
which become especially good for the
wide COMPASS $x$ region we deal with. 
Having in mind the surprisingly small \cite{dssv,EPJ2010} values of 
$\Delta_1 G$ and $\Delta_1 s$
it seems that we now became still more close to the 
minimal ``retro''
%``classical'' quantum-mechanical 
picture
of the proton spin puzzle, where
only  helicity PDFs and orbital moments of valence quarks 
%and their orbital moments 
compose the nucleon spin. 

Now we are waiting for the new COMPASS data with the planed $180\, GeV$ 
muon beam, which should allow to reach still smaller $x$ values (we expect for the
bottom boundary about $a=0.003$ instead of $a=0.004$) and, thereby, increase
the reliability of the presented results.

%Acknowledgement
  The authors are grateful to N.~Akopov, A.~Efremov, O.~Ivanov, 
 A.~Korzenev,  A.~Kotikov, V.~Krivokhizhin, 
 A.~Nagaytsev, A.~Olshevsky, 
G.~Piragino, G.~Pontecorvo,  I.~Savin, A.~Sidorov, O.~Teryaev, 
R.~Windmolders, and E.~Zemlyanichkina  
for fruitful discussions.

%*************************************************
\vskip 1cm

\begin{center}
        {\Large\bf Appendix\\
}
{\bf Combined analysis of  data coming from experiments with
 different binnings}
\end{center}
\renewcommand{\theequation}{A.\arabic{equation}}
\setcounter{equation}{0}

For brevity and readability we clarify our procedure  
of combined analysis on a simple example. 

Let us suppose that we would like to combine the data 
on the DIS virtual photon spin asymmetry
$$
A_1\simeq  g_1/F_1,
$$
coming from two independent
experiments ($exp1$ and $exp2$) with  different binnings. 
Of importance is that our final destination is the {\it integral} quantity -- 
truncated first moment -- 
\be
\label{integral}
\Gamma_{[a,b]} = \int_a^b dx\, g_1 =
\int_a^b dx\, A_1 F_1
\ee
of the structure function $g_1$, where the unpolarized structure 
function $F_1= F_2/2x(1+R)$ is considered as an already known\footnote{It is calculated 
in NLO QCD by using the known expression for DIS Wilson coefficients, 
some parametrization on unpolarized PDFs and the known parameterization on $R$.},
continuous function of $x$;
$a$ and $b$ are the lowest and highest $x$ boundaries available at least to one of the
considered experiments ($a=0.004$, $b=0.7$ for the combined analysis of COMPASS and HERMES
data we deal with).
Of importance is also that, just as in Ref. \cite{prd2006}, we apply the advanced approximation
of the integral by the sum over bins, 
which  reproduces much better the real integral value than the ``middle point'' approximation
(see discussion around Eq. (16) in Ref. \cite{prd2006}).
For the example considered here, it means that
we approximate by a constant within the bin only the measured asymmetry $A_1$:
$$
A_1(x)=\sum\nolimits_{i=1}^{N_{bins}}
A_1(\langle x_i\rangle)\, \theta(x-x_{i-1})\theta(x_i-x),
$$
where $A_1(\langle x_i\rangle)$ is the value of asymmetry 
in {\it i}-th bin, 
$x_0=a$, $x_{N_{bins}}=b$ and $\theta(x)$ is the usual step function,
while the known unpolarized input $F_1(x)$ remains a continuous function  
within each bin. Thus, 
instead of the rather crude approximation (``middle point'' approximation)\\ 
$
\Gamma_{[a,b]} \simeq \sum\nolimits_{i=1}^{N_{bins}} 
A_1(\langle x_i\rangle)\,F_1(\langle x_i\rangle)
$
to Eq. (\ref{integral}) we use the improved approximation 
\be
\label{aint}
\Gamma_{[a,b]} =\sum\nolimits_{i=1}^{N_{bins}}A_1(\langle x_i\rangle)\,
\int_{x_{i-1}}^{x_i}dx \, F_1(x).
\ee

Let us now consider the case where some bin $[\alpha, \beta]$ 
of $exp1$ is covered by 
 two bins  $[\alpha, \gamma]$ and $[\gamma, \beta]$
of  $exp2$. Then, to 
safely combine the respective data with Eqs. (\ref{stat_comb_value}), (\ref{stat_comb_error})
we apply the following procedure. First, we artificially divide bin  $[\alpha, \beta]$  
of $exp1$ into two bins  $[\alpha, \gamma]$, $[\gamma, \beta]$
and assign to each of the two new bins a
pseudo-measurement of the asymmetry $A_1$. These are $\tilde{A} \pm \delta\tilde{A}$        
and $\dtilde{A} \pm \delta\dtilde{A}$ for the artificial bins  
$[\alpha, \gamma]$ and $[\gamma, \beta]$, respectively, while the measured asymmetry 
in the initial real bin $[\alpha, \beta]$ of $exp1$ is just 
$A \pm \delta A$.
Then, we impose the natural requirement on our imaginary measurements:
\be
\label{our_comb_values}
\tilde{A} = \dtilde{A}=A. 
\ee
Now the pseudo-errors $\delta\tilde{A}, \delta\dtilde{A}$ should be properly adjusted.
We do it by imposing the necessary condition 
that the constructed system of pseudo-measurements must be
statistically equivalent to the system of real measurements within $exp1$.
Thus, a first relation connecting the errors $\delta A$ and $(\delta\tilde{A},\, 
\delta\dtilde{A})$ is  obtained by applying equation (\ref{stat_comb_error}):
\begin{equation}
\label{our_comb_syst1}
(\delta A)^2 = \frac{1}{1/(\delta \tilde{A})^2 + 1/(\delta \dtilde{A})^2}. 
\end{equation}
At the same time one can see  that due to Eq. (\ref{our_comb_values}) the second 
equation (\ref{stat_comb_value}) 
of the statistical addition just transforms into identity.

A second relation for the errors is obtained 
by taking into account that the requirement of statistical equivalence should also be satisfied
in the  calculation of the integral quantity $\Gamma_{[a,b]}$ we are interested in.
In the example considered here it means that  
the quantity (entering the sum in Eq. (\ref{aint}))
$\Gamma_{[\alpha,\beta]}|_{\text{I}} = A \int_{\alpha}^{\beta}dx\,F_1 $
and its error
$
\delta \Gamma_{[\alpha,\beta]}|_{\text{I}}   
$
should not change when
 the bin $[\alpha,\beta]$ of  $exp1$ is divided
into two pseudo-bins $[\alpha,\gamma]$ and $[\gamma,\beta]$, 
from  which the same quantity is derived: 
\be
\Gamma_{[\alpha, \beta]}|_{\text{II}} &=& \tilde{A} \int_{\alpha}^{\gamma} dx F_1 
+ \dtilde{A} \int_{\gamma}^{\beta} dx F_1
\pm \delta \Gamma_{[\alpha, \beta]}|_{\text{II}}.
\ee

The requirement of the central values equality 
\begin{equation}
\Gamma_{[\alpha, \beta]}|_{\text{I}} = \Gamma_{[\alpha, \beta]}|_{\text{II}}
\end{equation}
is satisfied automatically by virtue of Eq. (\ref{our_comb_values}),
while the requirement
\begin{equation}
\delta \Gamma_{[\alpha, \beta]}|_{\text{I}} = \delta \Gamma_{[\alpha, \beta]}|_{\text{II}},
\end{equation}
is written as
\be
\label{our_comb_syst2}
\left(\delta A \int_{\alpha}^{\beta} dx\, F_1\right)^2  
=\left(\delta\tilde A \int_{\alpha}^{\gamma} dx\, F_1\right)^2 +
\left(\delta \dtilde A \int_{\gamma}^{\beta} dx\, F_1\right)^2.
\ee
Solving the system (\ref{our_comb_syst1}), (\ref{our_comb_syst2})
one obtains
\be
\delta\tilde{A} = \delta A \sqrt{\frac{\int_{\alpha}^{\beta}  dx\, F_1}
{\int_{\alpha}^{\gamma} dx\, F_1}}, \quad
\delta\dtilde{A} = 
\delta A \sqrt{\frac{\int_{\alpha}^{\beta} dx\, F_1}
{\int_{\gamma}^{\beta}  dx\, F_1}}.
\ee

We operate  in the same  way every time when some bins of $exp1$ and $exp2$
do not coincide with each other and, after that, safely apply Eqs. (\ref{stat_comb_value}), 
(\ref{stat_comb_error}) to combine the data of both experiments.

It is easy to see that the generalization of the proposed procedure 
is straightforward in the case of 
difference SIDIS asymmetries (or  in the case of any other asymmetry 
we deal with).
For these asymmetries 
the role of the integral quantity $\Gamma$ in the above procedure is
played by the integral quantities ${\cal A}_{p}^{(n)}, {\cal A}_{d}^{(n)}$ (see (\ref{aint2})),
giving direct access to the {\it n}-th moments of valence PDFs we are interested in
(see (\ref{fmain2})).

%*************************************************

%%%%%%%%% Figures %%%%%%%%%%%%%%%%%%%%%%%%%%%%%%%%%%%%%%%%%%%%%%%%%%%%%%%%%%%%%%%%
\clearpage

\end{document}